# CH$_4$ reforming with CO$_2$ in a nanosecond pulsed discharge. The importance of the pulse sequence.

Cesare Montesano[a], Marzia Faedda[b,a], Luca Matteo Martini[a], Giorgio Dilecce[c,a], Paolo Tosi[a,c,ƒ]

[a] *Department of Physics, University of Trento, Italy*
[b] *Faculty of Fuel Technology AGH - University of Science and Technology, Kraków, Poland*
[c] *CNR, Istituto per la Scienza e Tecnologia dei Plasmi (ISTP), Italy*

[ƒ]Corresponding author e-mail address: paolo.tosi@unitn.it (P. Tosi).

**Abstract**

The plasma dry reforming reaction of methane with carbon dioxide is investigated in a nanosecond repetitively pulsed discharge, a type of plasma that offers some of the highest performance and non-equilibrium characteristics. The experiment's purpose was to examine the effect of varying the sequence of high-voltage pulses. We find that when successive pulses are closer than 500 μs, a memory-dominated regime gradually develops, which influences subsequent breakdown events. While reactant conversions increase with the plasma energy, both energy efficiency and conversions increase by shortening the inter-pulse time at the same plasma energy. This finding suggests that plasma power is not the only thing that matters to achieve better performance. How it is delivered can make a significant difference, in particular for CO$_2$, whose conversion doubles at the maximum energy for molecule investigated, 1.6 eV molecule$^{-1}$.

## 1. Introduction

Although oil and coal are the most used fuels, natural gas will play a growing role. In 2018, the global share of natural gas in the primary energy mix increased to 24%, with one of the most substantial growth rates for over 30 years. Natural gas is primarily methane, which has higher specific energy than other fuels and is also the cleanest burning fuel. Unfortunately, many gas reserves are in remote areas, which often prevents their exploitation due to the enormous transportation costs. The low energy density of gas makes its transport extremely disadvantageous compared to liquid energy vectors. This problem also plagues gas associated with oil deposits, which is frequently flared, posing severe environmental implications.

Methane emissions have increased in the past decade, with a substantial contribution from the agriculture, waste and fossil fuel sectors [1]. Since the global warming potential of CH$_4$ is much higher than that of CO$_2$, the problem arises of reducing atmospheric emissions of methane. In situ reforming to liquid fuels and other value-added products can be a valuable option for mitigating emissions and making more efficient energy transport. Already available technology for chemically "liquefying" CH$_4$ consists of methanol production from syngas produced by steam reforming CH$_4$ + H$_2$O → CO + 3H$_2$.

Among the various processes for methane reforming, the reaction with carbon dioxide, the so-called dry reforming (DR) CH$_4$ + CO$_2$ → 2CO + 2H$_2$, is desirable [2-3]. It uses one CO$_2$ molecule for each CH$_4$ molecule, producing syngas with a higher CO content suitable for producing synthetic hydrocarbons. Therefore, DR can help to reduce the carbon footprint of methane conversion. However, dry reforming is a very endothermic reaction ($\Delta H_{298K}$ = 247 kJ mol$^{-1}$) and still impractical at an industrial scale. Paradoxically, the high energy request that makes DR inefficient for thermal technology makes it an opportunity to store renewable electricity. Long-term storage is a fundamental issue for the massive exploitation of renewables.

The power-to-fuel concept is based on electricity use to promote endothermic reactions, thus converting electricity into chemical energy. The advantages are manifold: electricity from intermittent sources can be stored for long periods in a stable and dense form; methane and electricity can be converted into liquid fuels appropriate for those sectors where energy density is pivotal, e.g. hauling and aviation; The energy distribution can become more efficient and less costly.

Plasma discharges are especially suitable for energy storage [4,5]. Directly powered by renewable electricity, plasma reactors can use excess energy to produce value-added products from $CO_2$ that would otherwise be flowed in the atmosphere. Besides, plasma reactors can be switched on/off quickly, thus matching the variability of renewable energy sources. Finally, plasma discharges can be out from thermal equilibrium, with the electrons' kinetic energy much higher than that of the more massive particles. The vibrational temperature may also be higher than the gas temperature [6]. This peculiar mean-energy hierarchy can be usefully exploited to drive chemical reactions with minimised thermal energy losses because the input energy is primarily concentrated in the production of highly reactive precursors [7].

Different kinds of discharge have been used so far, with and without catalysts [8]. Unfortunately, it is not easy to compare efficiencies, given the lack of a unique definition. In a previous paper [9], we introduced the concept of Energy Conversion Efficiency, ECE, defined as the ratio between the low heat value of the products and the sum of the converted reactant ones with the energy E injected into the plasma

$$ECE = \frac{LHV_P}{LHV_R^{conv} + E}$$

In [9], we compared the performance of the nanosecond repetitively pulsed discharges (NRP) with other types of plasmas, finding that NRP discharge has higher ECE than Dielectric Barrier Discharges (DBD) and similar to Gliding Arcs (GA) and spark discharges. DBDs have also been used for investigating the synergy between plasma and catalysis [10]. Recently, conversions higher than 50% have been reported for $CH_4$ and $CO_2$, with efficiency up to 0.59 mmol $kJ^{-1}$ [11]. The last number translates in an ECE value of about 20% that, while being a valuable improvement for DBD, is significantly lower than ECE values (in the range 40-60%) obtained by NRP, GA, and spark discharges without catalysts [9].

Various authors have recently begun investigating the effects of changing the discharge's pulsing sequence [12-18]. We have shown that in an NRP, both the $CO_2$ dissociation and the energy efficiency increase by shortening the inter-pulse time $T_p$ for the same total energy [18]. Below 100 μs, subsequent pulses do not act independently; instead, they occur in a medium perturbed by the initial pulse. In such a situation, the discharge energy is dissipated at comparatively lower electron energy and higher electron density, which likely favours a more efficient splitting of $CO_2$.

Here, we investigate the dry reforming reaction in an NRP discharge by changing the pulsing scheme. We find that injecting the same energy using different voltage pulse sequences influences the dry reforming outputs remarkably. Both reactant conversions and ECE increase by shortening the inter-pulse time.

## 2. Experimental setup and methods

The experimental setup is shown in Fig. 1. The discharge operates close to atmospheric pressure (980 mbar) in a pin-to-pin configuration. The anode is a tungsten tube with an outer diameter of 3 mm and an inner diameter of 2 mm; the cathode is a tungsten rod with a 2 mm diameter. The gap is kept constant to 5 mm. The electrodes are held by brass cylinders inserted in a PTFE and Macor ceramic structure to ensure the sealing. The reactor chamber is made of stainless steel. The discharge is ignited using a generator (NPG 18/100k, Megaimpulse Ltd.) capable of producing high voltage pulses, with full width at half maximum (FWHM) of about ten ns and rise time < 4 ns. The trigger is provided by a waveform generator (32250A, Agilent Technologies Inc.). We used two different pulsing schemes shown in Fig. 2. The first one, called continuous mode, consists of a constant repetition of equally spaced pulses, with frequency in the range 450 – 1200 Hz. The second scheme, called burst mode, is made of a few much closer pulses (frequency in the range 2 - 50 kHz) grouped into repetitive bursts whose frequency is 300 Hz.

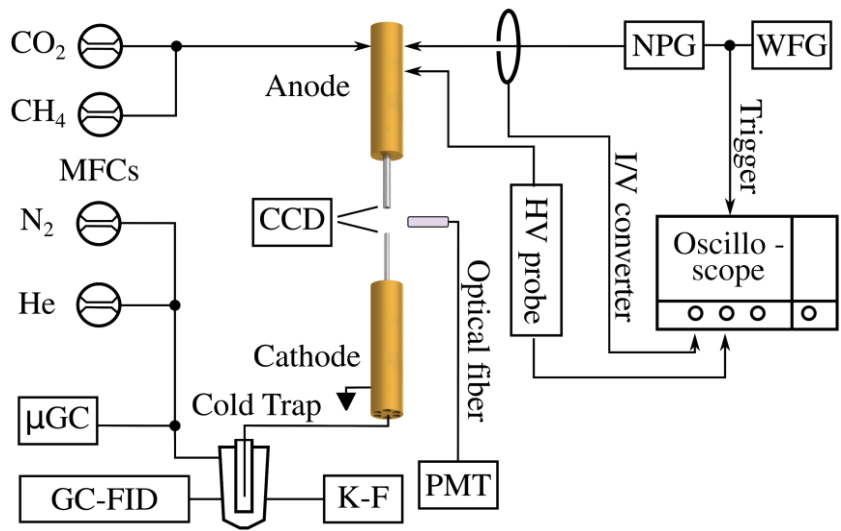

Fig. 1: Experimental setup. MFCs: mass flow controllers; K-F: Karl-Fisher volumetric titrator; PMT: photomultiplier tube; HV probe: high voltage probe; NPG: nanosecond pulse generator; WFG: waveform generator; (μ)GC: (micro) gas-chromatograph; FID: flame ionization detector; CCD: CCD camera.

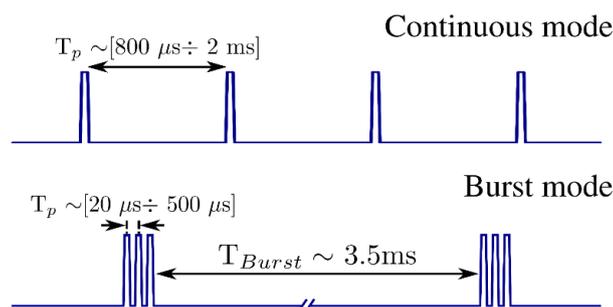

Fig. 2: Schematics of the pulsing schemes. Continuous mode (top), burst mode (bottom).

Current and voltage are measured by a I/V converter (CT-D-1.0, Magnelab, bandwidth = 200 Hz-500 MHz) and a high-voltage probe (P6015A Tektronix, bandwidth = 75 MHz), respectively. A digital oscilloscope (HD9104 Teledyne Lecroy, bandwidth = 1 GHz) acquires the signals with a sampling rate of 20 GS/s.

Measured voltage/current values are affected by a spurious time-delay τ introduced by the measurement setup. This delay must be considered to avoid substantial mistakes in estimating the plasma energy. τ is evaluated by filling the reactor with Freon-113 and acquiring voltage and current signals. Since the discharge is off, the power is purely reactive. We retrieved τ by fitting the measured current with the displacement current obtained by differentiating V(t). One gets both the delay (about 0.5 ns) and the reactor's capacitance (approximately 20 pF). The instantaneous power P(t) is obtained by the product V(t)·I(t). The pulse energy is the time integral of P(t) throughout the pulse.

A phototube (H10721-210, Hamamatsu) detects the discharge's optical emission, while a CCD camera (DMK 37BUX273, The Imaging Source) is used to image the discharge. Two mass-flow controllers (1179A, MKS inst.) regulate the reactant gas flux Φ. It is kept constant to 150 sccm during the experiment, with the $CO_2$:$CH_4$ ratio equal to 1. $N_2$ and He are added after the discharge and before the gas-chromatograph (GC), serving as dilution and internal standard (IS). IS is necessary to account for the moles' variation due to chemical reactions such that the outgoing gas flow from the discharge is different from the entering one. In other words, the conversion cannot be calculated by merely considering the peak area in the chromatogram, which is proportional to the molar fraction. By adding a constant amount of internal standard, one can estimate the factor $\xi = \frac{IS^{off}}{IS^{on}}$, the ratio between IS signals with the discharge off and on, needed to detect changes in the mole number. In-line analysis of the effluent mixture is carried out by a micro gas-chromatograph (3000 μGC, Agilent Technologies Inc.). The latter is equipped with two columns: a MolSieve 5A with back-flush to detect He, $H_2$, $N_2$, $CH_4$ and CO; a Plot U for $CO_2$, C2-hydrocarbons and propane. Calibration of the instruments with the external-standard method allows determining the (thermal conductivity TCD) detector's response factor for each compound of interest.

Light hydrocarbons (C3-C4) are quantified by a GC (Trace GC Ultra, Finningham) equipped with a flame ionization detector (FID) and a Plot Q column (HP-PLOT Q, Agilent J&W). The following temperature program was used: 50 °C, 5 min; 260 °C, 15 °C min$^{-1}$; 260 °C, 15 min. The injector is heated at 270 °C. The carrier gas is He (2.1 ml min$^{-1}$), split ratio 1/10. Propane is used as an internal standard. The FID response factor for C3-C4 hydrocarbons relative to propane is calculated based on the effective carbon number concept [19]. The error of this procedure is about 10%.

We also used a T-stabilised cold trap (-30°C) to condensate high vapour pressure species. The condensed species were dissolved in 10 ml of acetonitrile with the addition of anisole (4.0 mg ml$^{-1}$ in acetonitrile) as the internal standard. Samples (1 ul) were analysed by the same GC system (temperature program: from 100 °C to 260 °C, 15 °C min$^{-1}$; 260 °C, 20 min; splitless injection). FID response for methanol and C1-C4 carboxylic acids is obtained from a calibration method using standard solutions of analytical grade pure compounds in acetonitrile. The water condensed in the cold trap is quantified using the Karl-Fisher technique (Volumetric KF Titrator, Metter Toledo).

The following quantities are defined for characterising the plasma process. The conversion of the species X is:

$$C_X = \frac{n_X^{conv}}{n_X^{in}}.$$

Where $n_X^{conv} = n_X^{in} - n_X^{out}$ and $n_X^{in/out}$ indicates the number of moles of the input/output species X.

The product selectivities are

$$S_{H_2} = \frac{n_{H_2}}{2n_{CH_4}^{conv}} \times 100,$$

$$S_{CO}^C = \frac{n_{CO}}{n_{CH_4}^{conv} + n_{CO_2}^{conv}} \times 100,$$

$$S_{CO}^O = \frac{n_{CO}}{2n_{CO_2}^{conv}} \times 100,$$

$$S_{C_xH_y}^H = \frac{yn_{C_xH_y}}{4n_{CH_4}^{conv}} \times 100,$$

$$S_{C_xH_y}^C = \frac{xn_{C_xH_y}}{n_{CH_4}^{conv} + n_{CO_2}^{conv}} \times 100.$$

All the above quantities are evaluated as a function of the specific energy input (SEI):

$$SEI(kJ\ dm^{-3}) = \frac{P(kW)}{\Phi(dm^3\ s^{-1})},$$

$$SEI(eV\ molecule^{-1}) = SEI(kJ\ dm^{-3}) \times \frac{6.24 \cdot 10^{21}(eV\ kJ^{-1}) \times 22.4\ (dm^3\ mol^{-1})}{6.022 \cdot 10^{23}(molecule\ mol^{-1})}.$$

In the continuous mode, the power P is calculated as the product between the pulse energy and the repetition frequency; by changing the latter, one can vary the SEI. In the burst mode, the power is obtained by multiplying the burst repetition frequency times the sum of each pulse energy. The SEI is changed by varying the number of pulses.

## 3. Results and discussion

The two pulsing schemes present the same general features discussed in [18]. In the continuous mode, the pulses are separated by inter-pulse times longer than 800 µs and have a similar aspect. This similarity is explained by the extended distance between the pulses, which allows the system to relax to the same initial conditions. The pulses show a first breakdown region wherein a high voltage is required to initiate the discharge. Later, multiple strokes, characterised by a lower voltage, are ignited by the reflected power travelling back and forth along the output cable. Other sequences of discharge events appear around 0.5 and 2 µs. The origin of these voltage rises is unknown. We believe they are linked to a retrigger of the power supply. While a small amount of energy is dissipated in these bumps, some light is still observed, indicating a discharge ignition.

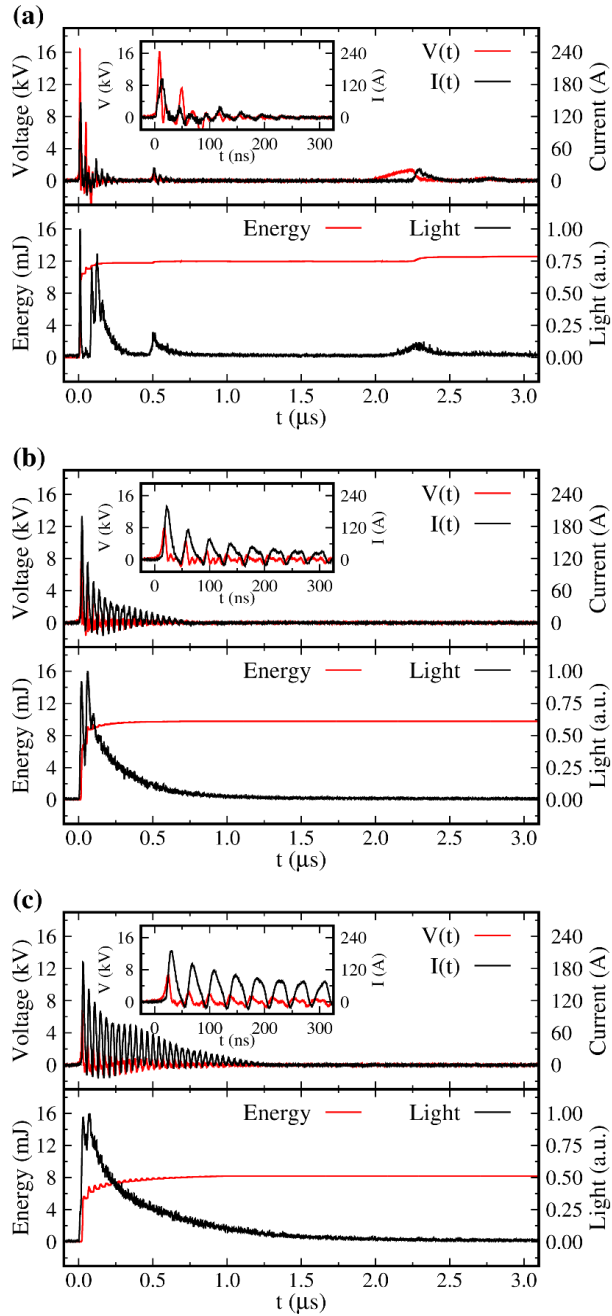

Fig. 3: The first (a), second (b), and third (c) pulse of a 25 kHz burst, separated from each other by 40 µs.

When grouped in a burst at short intervals, pulses are pretty different from each other. Fig. 3 shows the current and voltage characteristics of a burst with an inter-pulse time of 40 µs. The first pulse, shown in Fig. 3a, looks like those of the continuous mode. The successive are dissimilar, as shown in Fig. 3b and 3c. Here, voltage and current oscillate more regularly. A lower voltage and higher current characterise those pulses. All the reflected power peaks can reignite the discharge, as indicated by the intense, long-lasting light emission. The bump at 2 µs is absent. The difference between the first and second pulse begins to be discernible already in the 2 kHz burst, corresponding to 500 µs distance, see Fig. 4. In the second pulse, voltage and current show more regular oscillations, the bump of the voltage around 2 µs is still present, but its current is smaller than in the first pulse. The reason likely lies in the different gas composition encountered by the second pulse, which modifies the discharge impedance.

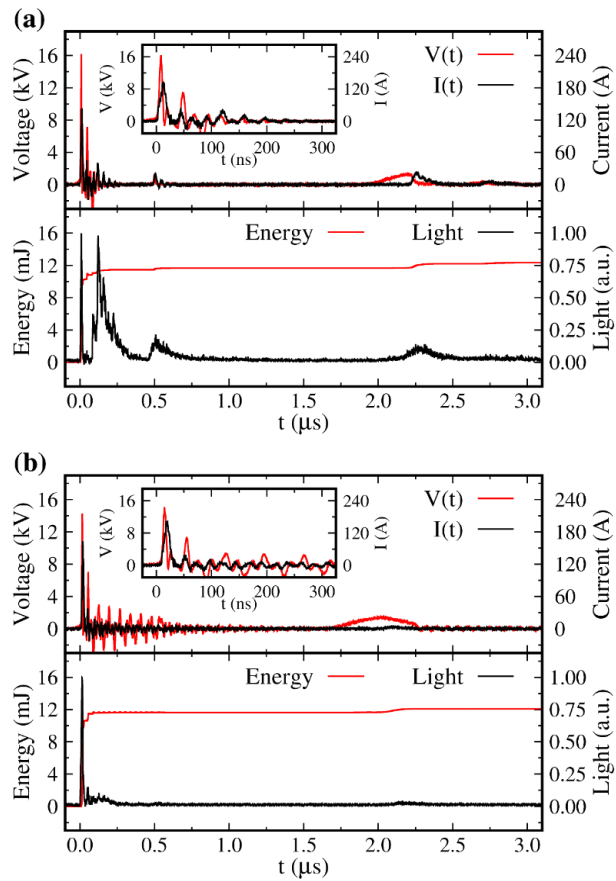

Fig. 4: The first (a) and second pulses (b) of a 2 kHz burst ($T_P$=500 µs).

To better quantify the discharge evolution, we compare the total charge (Q) and breakdown voltage ($V^{bd}$) of the first two pulses in Fig. 5 and 6. The charge is obtained by integrating the current over time. The breakdown voltage is calculated by looking at where the current deviates from the displacement current. The greater charge of the first pulse over the second at large $T_p$, which seemingly contrasts with the almost equal total energy, is due to the current's contribution at 2 µs.

As expected, the charge, energy and the breakdown voltage of the first pulse do not change with $T_p$. On the contrary, the second pulse on decreasing $T_p$ shows a progressive charge increment and breakdown voltage reduction. A memory effect gradually develops on shortening the inter-pulse distance.

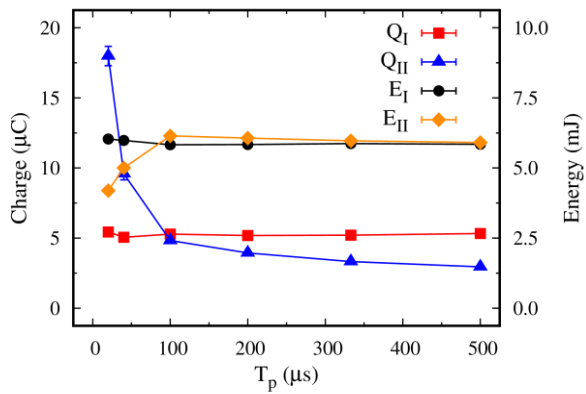

Fig. 5: The graph shows the total charge (Q) and energy (E) of the first two pulses in different bursts. The subscripts I and II refer to the first and second pulse, respectively.

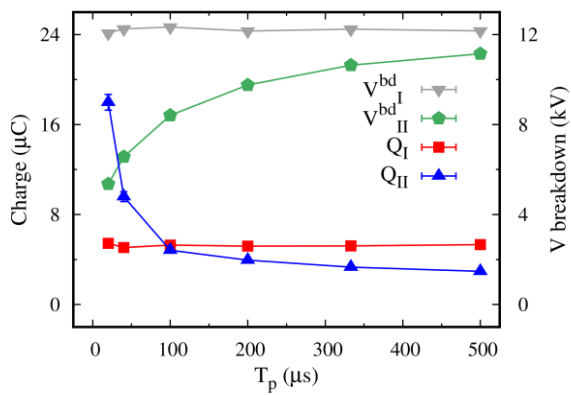

Fig. 6: The breakdown voltage for the first two pulses ($V^{bd}_I$ and $V^{bd}_{II}$) is shown together with the charge ($Q_I$ and $Q_{II}$) deposited by each pulse.

The memory effect is due to a modification of the discharge conditions. Each discharge triggers a gas expansion due to increased temperature (result of the Joule heating and exothermic back reactions) and mole number caused by the conversion of the reactants [7]. The overall effect of changes in gas temperature, pressure and composition, can lead to a lower breakdown voltage and a difference in the discharge channel development (different load impedance). The memory effect becomes dominant below 100 µs, where the current changes drastically, as shown in Fig. 3b,c. This regime change corresponds to a $T_p$ threshold (around 40-100 µs) below which all the subsequent discharge pulses follow the first pulse's path. This behaviour is evident in Fig. 7, where the images of a three-pulse burst are shown. The images are acquired with an exposure time large enough to catch all burst's three pulses. From 100 µs up to 500 µs, each pulse is spatially independent; for shorter inter-pulse times, 20 and 40 µs, the second and third discharges appear to follow the same path. In Fig. 7, the I/V characteristics of the second pulse in the burst are also shown alongside the corresponding image.

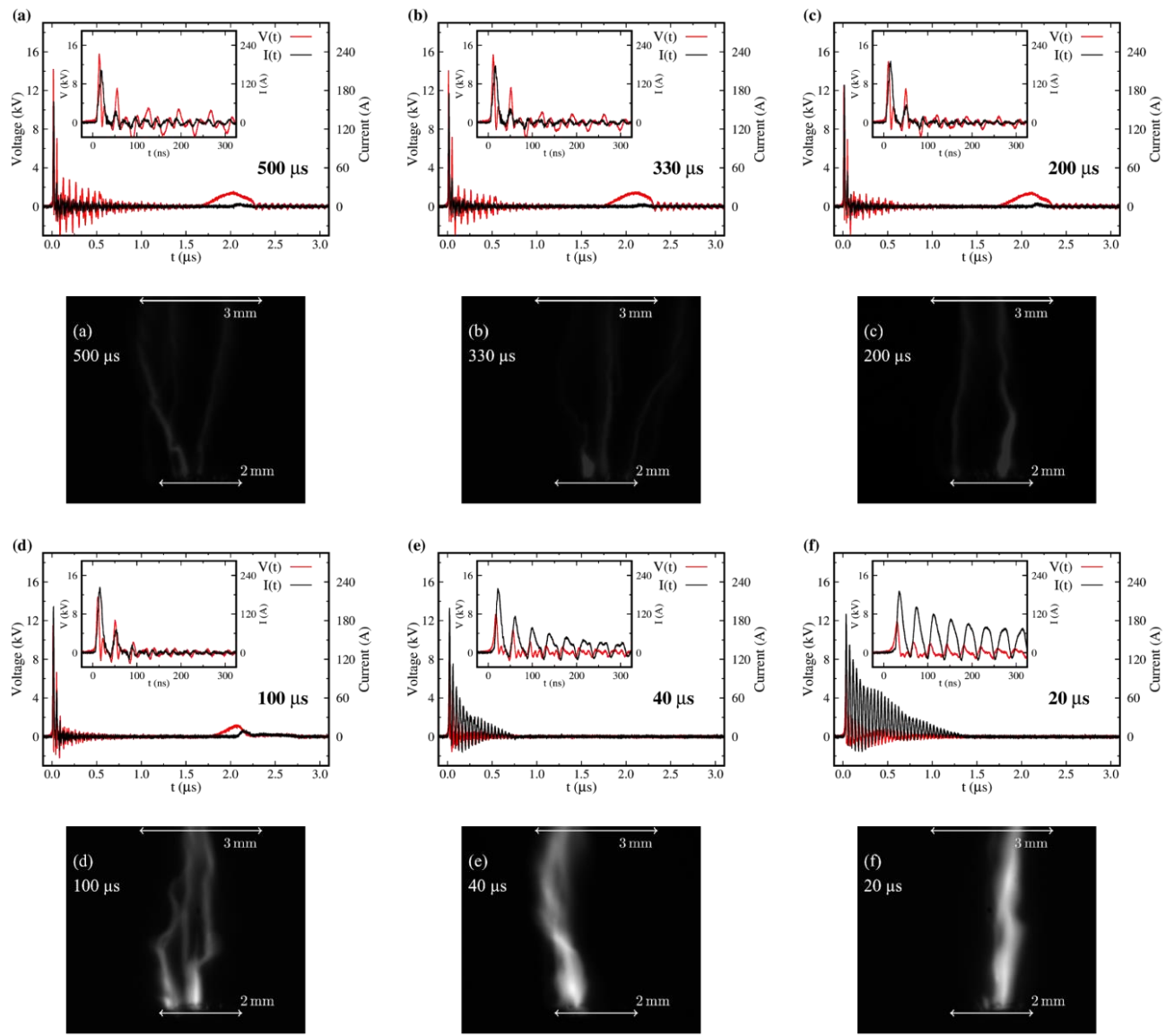

Fig. 7: (First and third rows) I/V characteristics of the second pulses in the bursts with inter-pulse times ($T_p$) ranging from 500 µs to 20 µs. (Second and fourth rows) Images of a three-pulse burst discharge at various inter-pulse times showing the path of different pulses. The CCD is gated synchronously to the discharge, with an exposure time as ample as to capture all pulses in a unique acquisition. The image intensity scale of each panel is arbitrary. Electrodes position and diameter are indicated by white arrows: the anode is in the top part of the pictures (outer diameter of 3 mm); the cathode is in the bottom part (diameter of 2 mm).

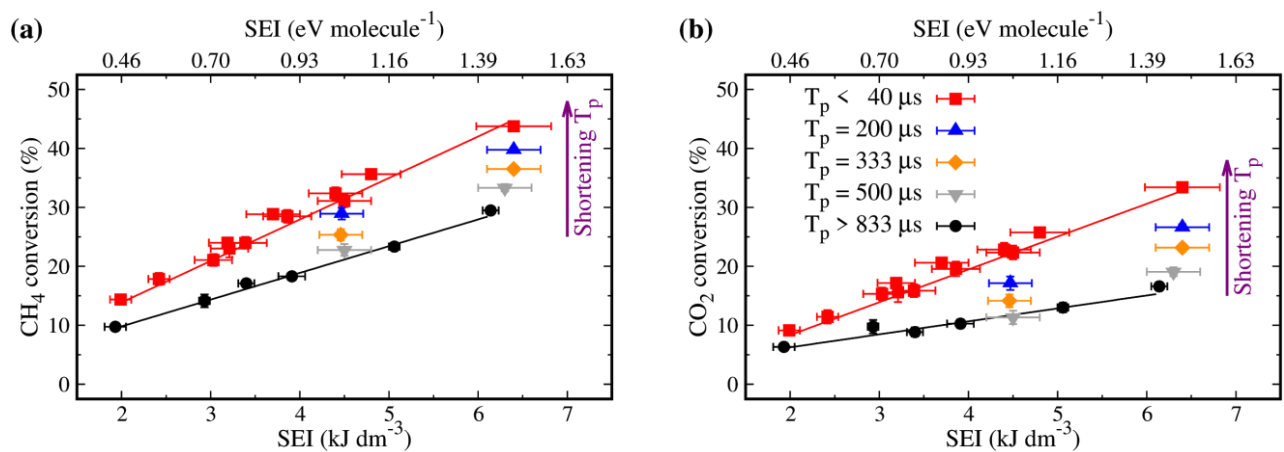

Fig. 8: Reactant conversion as a function of SEI for various interpulse intervals ($T_p$). Error bars account for statistical errors. The legend is the same for both plots, and it is reported in (b).

Fig. 8 shows the conversions of methane and carbon dioxide as a function of the SEI for different inter-pulse intervals. At each SEI, the dissociation values increase by shortening the inter-pulse time. The highest dissociations of $CO_2$ and $CH_4$ are achieved using a repeating sequence of a few very close (≤ 40 μs) pulses rather than a continuous train of relatively distant (0.8 - 2 ms) pulses. The systematic error on the chromatographic measurements accounts for about 5%; for the SEI values, it accounts for about 7%. The product selectivity is also clearly affected by the pulsing scheme. See Fig. 9 to compare values obtained in the continuous mode with those of the shortest $T_p$ (≤ 40 μs) burst.

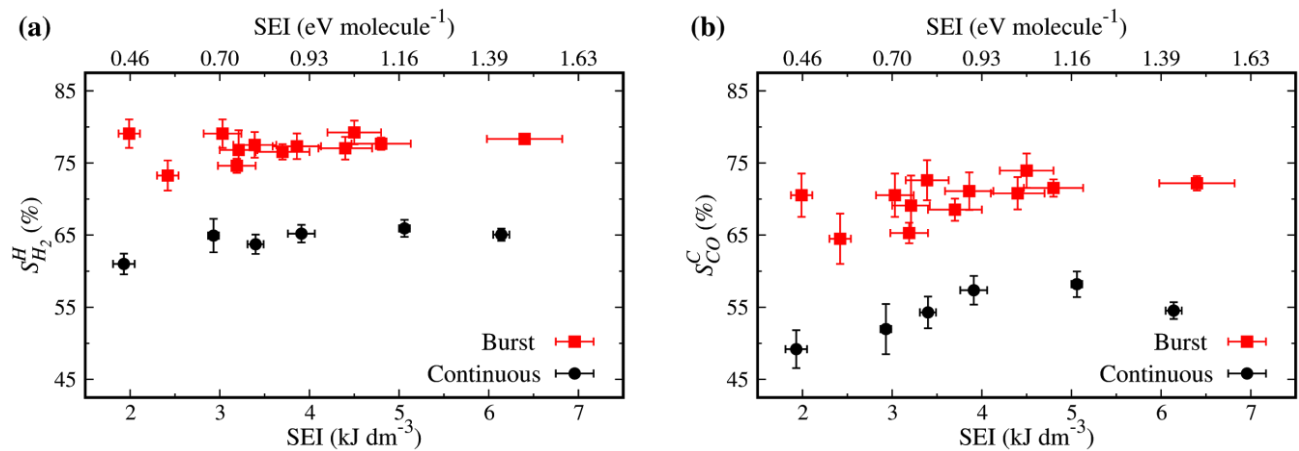

Fig. 9: (a) Selectivity of $H_2$ with respect to H, (b) Selectivity of CO with respect to C. Error bars account for statistical errors.

A higher syngas (CO and $H_2$) selectivity is obtained with very close pulses (burst mode). The situation is reversed for the selectivity towards C2-hydrocarbons production, which is higher in the continuous mode, see Fig. 10. For C3-C4 hydrocarbons, the difference is not so clear. All selectivity values appear to change very little in the investigated range of SEI, apart from $S_{CO}^C$, which shows a broad maximum at 4-5 kJ dm$^{-3}$. In the burst mode, the values are, on average, $S_{H2}^H = 77\%$, $S_{C2}^H = 12\%$, $S_{C3C4}^H = 3\%$. In the continuous mode, $S_{CO}^C = 54\%$, $S_{C2}^C = 28\%$, $S_{C3C4}^C = 7\%$, $S_{H2}^H = 64\%$, $S_{C2}^H = 19\%$, $S_{C3C4}^H = 4\%$. Within the family of C2 hydrocarbons, the acetylene selectivity is larger in the burst mode.

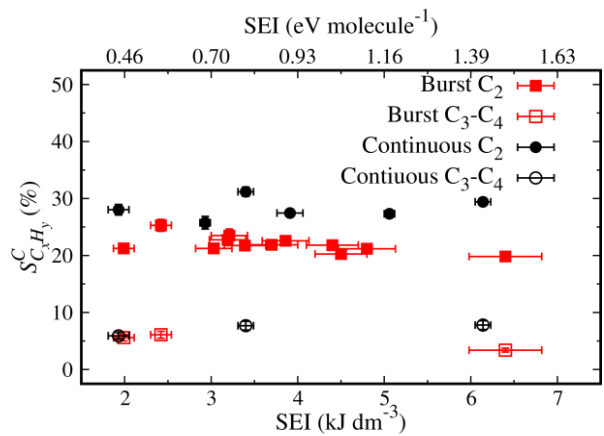

Fig. 10: Selectivity of hydrocarbons with respect to C. Error bars account for statistical errors.

This finding might be explained by the different temperatures reached by shortening the inter-pulse time. As discussed in [17, 20-21], the production of $C_2H_2$ is promoted at temperatures above 1500 K against $C_2H_4$ and $C_2H_6$. A further increment of the gas temperature moves the balance toward the acetylene rather than the other C2 hydrocarbons, matching the different discharge conditions in the continuous and burst mode. The liquid condensed in the cold trap is mainly water (the estimated upper limit for $S_{water}^H$ is 10%). Methanol and carboxylic acids are detected in traces (account for less than 5% of the total liquids collected). Overall, the calculated mass balance is around 100%. Minor deviations are due to non-quantified chemicals such as other alcohols and carbon powder.

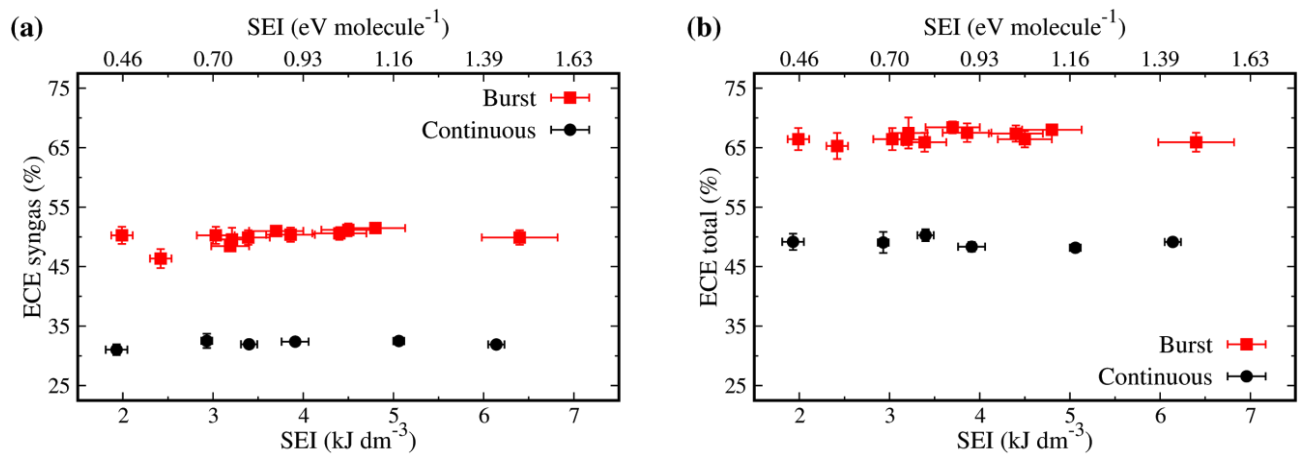

Fig. 11: (a) ECE for syngas production; (b) ECE for syngas plus hydrocarbons production. Error bars account for statistical errors.

The discharge efficiency is assessed separately by considering only the syngas production or accounting also for hydrocarbon production. In Fig. 11, we compare ECE values for the continuous and the burst mode (Tp < 40 μs). The burst offers a total ECE of around 65%, while it is 50% in the continuous mode. A significant difference also appears when only syngas is considered.

The conversion correlates well with the discharge parameters reported in Fig. 5 and 6. All these quantities are plotted together in Fig. 12. The steady increase in conversion on reducing $T_p$ corresponds to a charge increase in the discharge and decrease of the breakdown voltage. We point out that conditions of lower voltage, i.e. lower electron mean energy, and higher current, i.e. higher charge density, favour the vibrational excitation of molecules and then perhaps enhance the role of the vibrational dissociation mechanism.

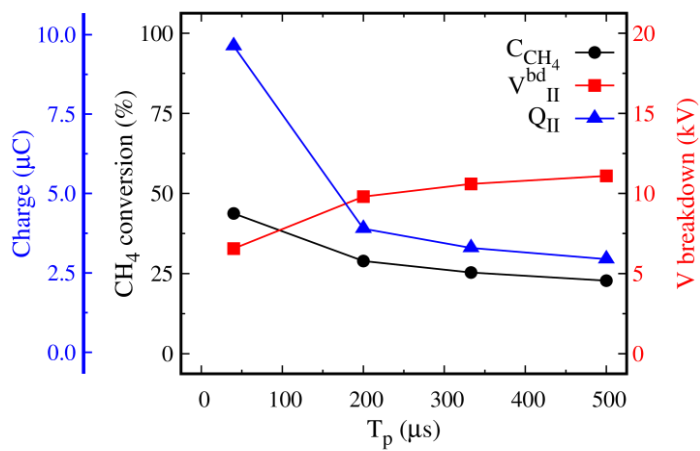

Fig. 12: $CH_4$ conversion ($C_{CH_4}$, black), breakdown voltage of the second pulse ($V^{bd}_{II}$, red), and Charge ($Q_{II}$, purple) as a function of $T_P$.

Maximum rate coefficients for electron-impact vibrational excitation of $CO_2$ are obtained at electron temperatures around 1-2 eV [22]. Spectroscopic measurements in a pure $CO_2$ discharge [23] show that in the second part of the first pulse (after the initial breakdown region) and the successive pulses, a spark regime is established, characterised by an electron temperature around 2 eV and electron density peak values of the order of $10^{18}$ cm$^{-3}$.

On the contrary, the initial breakdown phase (the first voltage/current peak) features lower electron density, unmeasurable but certainly less than $10^{17}$ cm$^{-3}$, and higher - although non-measurable - electron mean energy. An enhanced vibrational excitation of $CO_2$ may interpret the burst mode's better performances, although quantitative modelling still lacks. In addition, a thermal effect contributes to the burst mode's efficiency, in particular for $CH_4$ [21, 24]. Time-resolved gas temperature measurements in these conditions must rely on spectroscopic methods and are not easy. In [23], we attempted measures by added $N_2$ trace emissions, finding temperatures in the surroundings of 2500 K in the first pulse, with a slight increase in successive pulses challenging to quantify. In the present gas mixture, further thorough analysis of the emission spectra must be done to find a reliable measurement method beyond this paper's scope. Also, by lowering the interpulse delay might result in successive pulses igniting a gas where a higher density of active species can be formed, eventually being beneficial for converting reactants [12, 24].

## 4. Conclusions

This paper reports results for the dry reforming reaction obtained in an NRP discharge for different inter-pulse times. We observe that both reactant conversion and energy efficiency increase by shortening the time interval between successive discharge pulses, with saturation below 40 µs. At the maximum explored SEI, around 6 kJ dm-3, the $CO_2$ conversion doubles, and that of $CH_4$ increases by almost 50% by shortening the pulse interval from $T_p > 833$ µs to $T_p \leq 40$ µs. The energy efficiency conversion ECE goes from 50% to about 65%. The product selectivity is also dependent on the inter-pulse time, with closer HV pulses bolstering CO and $H_2$ and more spaced pulses that favour the hydrocarbon production. We suggest that the increased performance observed by shortening the pulse separation is due to the progressive modification of the discharge conditions, both for the gas composition, temperature and load impedance. Closer pulses are coupled in a post-discharge medium not fully relaxed to the initial conditions. Pulses following the first are characterised by a comparatively lower breakdown voltage and a larger current. A significant amount of low-energy electrons is known to favour a vibrationally enhanced dissociation mechanism for $CO_2$. However, its role cannot be evidenced by measurements of the final mixture composition only. In particular, what happens when the distinct discharge paths collapse into a single channel is not clear. The substantial increase in the total charge does not correspond to an equally sharp rise of dissociation. In principle, the single-channel condition should boost the vibrational mechanism, but this is likely achieved in a volume smaller than that covered by the independent paths. Or even a partial saturation of the dissociation may occur. Hopefully, advanced plasma diagnostics [23,25-27] and future modelling efforts [7, 28] will help understand the complex interplay of discharge physics and plasma chemistry in such non-equilibrium discharges.

**Acknowledgements**

*This project has received funding from the European Union's Horizon 2020 research and innovation programme under the Marie Skłodowska-Curie grant agreement No. 813393.*